\documentclass[draft]{agujournal2019}
\usepackage{url} 
\usepackage{lineno}
\usepackage[inline]{trackchanges} 
\usepackage{soul}
\draftfalse

\journalname{Geophysical Research Letters}

\begin{document}

%
%


\title{Investigating the Dependence of Normalized Reconnection Rate on Upstream Plasma Parameters}

%
%




\authors{S. V. Heuer\affil{1}, K. J. Genestreti\affil{2}, Y.-H Liu \affil{3}, J. R. Shuster\affil{1}, X. Li \affil{4}, R. B. Torbert \affil{1,2}, J. L. Burch\affil{5}}

\affiliation{1}{University of New Hampshire, Durham, NH, 03824, USA}
\affiliation{2}{Southwest Research Institute, Durham, NH, 03824, USA}
\affiliation{3}{Dartmouth College, Hanover, NH, 03750, USA}
\affiliation{4}{Los Alamos National Laboratory, Los Alamos, NM, 87545, USA}
\affiliation{5}{Southwest Research Institute, San Antonio, TX, 78238, USA}





\correspondingauthor{S. V. Heuer}{svbheuer@gmail.com}



\begin{keypoints}
\item We perform a multi-event study of normalized reconnection rate to test for dependence on a range of upstream plasma conditions. 
\item We find no correlation between normalized reconnection rate with any upstream plasma conditions investigated, suggesting it is constant for typical magnetospheric conditions. 
\item This result may be significant in predicting the terrestrial effects of space weather events by providing insight into the efficiency of solar wind-magnetospheric coupling.
\end{keypoints}

%
%

%
%


\begin{abstract}
We present the results of a multi-event study of the normalized reconnection rate integrating events spanning the three primary regimes of reconnection observed by the Magnetospheric Multiscale (MMS) mission. We utilize a new method for determining the normalized reconnection rate with fewer sources of uncertainty by estimating the current sheet aspect ratio with magnetic field gradients, which are very well measured by MMS. This method is time-dependent and also captures spatiotemporal variation in the current sheet aspect ratio. After demonstrating this technique is valid in the guide field and asymmetric regimes of reconnection, we investigate any relationships between the normalized rate, aspect ratio, and spatiotemporal aspect ratio variability on the guide field, spatiotemporal variability of the upstream magnetic field, and upstream magnetic field and density asymmetry. We find no dependence of reconnection rate on upstream conditions, with the only statistically significant correlations between the spatiotemporal variability in the aspect ratio and the spatiotemporal angular variability in the upstream reconnecting component of the magnetic field. This result is consistent with previous work that found the spatiotemporal ``patchiness'' of energy conversion within the electron diffusion region is correlated with the same measure of spatiotemporal variability of the inflow magnetic field used here. Additionally, this analysis suggests that under typical magnetospheric conditions for density and magnetic field asymmetry with steady upstream magnetic field, the normalized reconnection rate is constant with a slight increase for higher guide fields, which may be significant in predicting the terrestrial effects of space weather by providing insight into the efficiency of solar wind-magnetospheric coupling. 
\end{abstract}

\section*{Plain Language Summary}
Magnetic reconnection is a fundamental process in plasmas which occurs throughout the universe. It can redistribute energy over vast scales, and accelerate and heat nearby particles. The rate at which reconnection occurs, known as the reconnection rate, is one of the critical parameters to understanding reconnection. In this Letter we perform an analysis of many reconnection events occurring with a wide range of plasma conditions to see whether the rate changes. We find no dependence of reconnection rate on any of the parameters investigated, although we do observe a correlation between spatial or temporal stability of the current sheet and angular variability of the upstream magnetic field. These results may be significant for understanding how well energy enters the magnetosphere from space weather events.

%
%

%


%
%
%
%

\section{Introduction}

\subsection{Background}

Magnetic fields in astrophysical plasmas have been observed to store very large amounts of energy, which can be released explosively through a process known as magnetic reconnection \cite{Zweibel_Yamada_2009}. Magnetic reconnection is a topological reconfiguration of the magnetic field which converts magnetic energy into particle kinetic energy through the generation of intense electric fields. The reconnection process occurs within a kinetic-scale region known as the diffusion region, where plasma is decoupled from the magnetic field \cite{Hesse2011}. While recent observations in space and laboratory have found reconnection occurring with only de-magnetized electrons \cite{Phan2018,Greess2021,PeiyunShi2023}, the diffusion region typically has a multi-scale structure with ions demagnetized in the ion diffusion region (IDR) at the ion inertial length with an embedded electron diffusion region (EDR) \cite{Vasyliunas1975,Hesse2011}. One of the most critical parameters for understanding the state of reconnection is the reconnection rate \cite{Comisso_Bhattacharjee_2016, Cassak_Liu_Shay_2017}, which has been the focus of many laboratory, spacecraft, and simulation studies across many plasma regimes. When comparing reconnection across different plasma regimes, it is common to consider the normalized reconnection rate, which is defined $V_{in}/V_{out}$ where $V_{in}$ is the inflow speed and $V_{out}$ is the outflow speed, typically taken to be the Alfven speed immediately upstream of the associated diffusion region \cite{burch2020_rxRate}. The normalized rate is also frequently written in terms of the tangential reconnection electric field normalized by the inflow Alfven velocity and reconnecting component of the magnetic field.

The normalized reconnection rate is commonly assumed to be of order 0.1 \cite{Cassak_Liu_Shay_2017} and is most often studied for the simplest case of anti-parallel, symmetric reconnection. The normalized reconnection rate, particularly whether it varies with the addition of a guide field or asymmetry in density or magnetic field, is of particular interest in the terrestrial magnetosphere where the reconnection rate is a key parameter in setting the efficiency of solar wind magnetospheric coupling and the rate of mass and energy transport in the magnetosphere \cite{McPherron1979,Baker1996}. Despite the importance of possible variation in the normalized reconnection rate there are conflicting results in the literature on possible correlations. Simulations have suggested that variations in reconnection rate up to a factor of two over the range of guide fields typically found in the magnetosphere \cite{Huba2005}, while others have indicated that there is no dependence at all \cite{Liu_2014_dispersive_waves,Pritchett2001}. In contrast, laboratory experiments have found that increasing guide field can decrease the normalized reconnection rate by almost an order of magnitude \cite{Tharp2013,Stechow2018}, while in situ spacecraft multi-event studies have found either no clear dependence of the rate on the guide field \cite{LJ_Chen_2017,Pritchard2023,ShangWang2015}, or a inverse dependence \cite{FuselierAndLewis2011}. 

This lack of consensus is partially because accurately measuring the normalized reconnection rate with spacecraft has many difficulties. The normalized reconnection rate is typically calculated using the inflow speed, normal magnetic field, and tangential reconnection electric field, which are typically the smallest components of their associated vectors \cite{FuselierAndLewis2011}. This makes the resulting rate very sensitive to errors associated with the determination of local coordinate system and velocity frame, which makes it difficult to determine any significant correlations between the normalized reconnection rate and local and background plasma conditions \cite{genestreti2018}. In addition to these measurement uncertainties most spacecraft missions have been unable to resolve the electron-scale physics within the reconnection region, further increasing the difficulty of studying the relation between kinetic-scale physics and the fluid-scale parameters thought to control the normalized reconnection rate \cite{Liu_2017,Liu_2022}. However, NASA’s Magnetospheric Multiscale (MMS) mission has made possible observations of both the macro-scale fluid quantities and the electron-scale physics within the diffusion region with unprecedented spatial and temporal resolution \cite{mms_burch2016}. 

\section{Outline of This Study}

In this Letter, we utilize the unique measurement capabilities of the MMS mission to analyze previously MMS identified EDR or near-EDR events to test the dependence of the normalized rate of reconnection on the guide field and density and magnetic field asymmetry. For this multi-event study we apply a new technique which leverages the unique capabilities of the MMS mission to determine the EDR aspect ratio and normalized reconnection rate with fewer sources of uncertainty than previous methods \cite{Heuer_2022}. This time-dependent method also captures spatiotemporal variability in the current sheet aspect ratio. While we find small variations in the normalized reconnection rate between events for all parameters, there are no consistent trends and the only statistically significant correlations are between the spatiotemporal variability in aspect ratio and spatiotemporal variability in the upstream reconnecting component of the magnetic field. 

Our approach to calculating the normalized reconnection rate for our multi-event study utilizes the aspect ratio of the diffusion region. While the aspect ratio of the diffusion region has been determined in situ using spacecraft in the past, the methods applied in past studies have all required assumptions about spacecraft trajectory, precise knowledge of velocity frame, and approximations for upstream parameters \cite{RNakamura2019,torbert_science2018, ShangWang2015}. In this Letter we find the diffusion region aspect ratio without any of these limitations by writing the aspect ratio $\delta/\ell$ in terms of the local magnetic field gradients \cite{Heuer_2022} 
\begin{equation}\label{eqn:aspect_ratio}
    \frac{\delta}{\ell} \simeq \bigg[ \bigg(\frac{\partial B_N}{\partial L} \bigg)^2 / \bigg(\frac{\partial B_L}{\partial N} \bigg)^2 \bigg]^{\frac{1}{4}}
\end{equation}

Where $L$ is the direction of the reconnecting component of the magnetic field and $N$ is the current sheet normal direction ($M$, which is not included in Equation (1) completes the right-handed coordinate system shown in Figure 1). The aspect ratio of the diffusion region $\delta/\ell$ is equivalent to $\tan(\phi)$, where $\phi$ is the opening angle, defined as the angle between the separatrix and the $L$-axis (see Figure 1). If we assume incompressibility and quasi-two-dimensonality, then the aspect ratio $\delta/\ell$ is equivalent to the normalized reconnection rate $V_{in}/V_{out}$. 
The velocities are typically evaluated at the upstream ($V_{in,e}$) and downstream ($V_{out,e} \simeq V_{Ae}$) boundaries of the associated electron-scale diffusion region (Figure 1) \cite{burch2020_rxRate,burch2022_edr_inflow_tail} although it is also common to use the comparable asymptotic quantities $V_{in,0}$ and $V_{out,0}\simeq V_{Ai0}$ \cite{FuselierAndLewis2011,ShangWang2015}. Recent theoretical and modeling advances have found that achieving the open outflow geometry necessary for fast reconnection requires cross-scale coupling of the kinetic-scale diffusion region with the asymptotic parameters to account for larger fluid-scale stresses \cite{Liu_2017}. Assuming steady-state reconnection, the normalized rate can be written explicitly as a function of aspect ratio 

\begin{figure}
\includegraphics{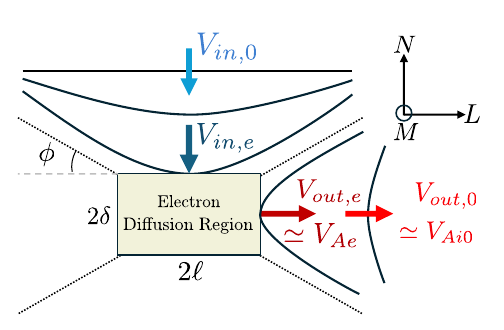}
\caption{\label{fig:1} Cartoon showing the geometry of symmetric reconnection with important quantities labeled. $2\delta$ and $2\ell$ are the height and width of the EDR, respectively, while an $e$ subscript indicates an electron-scale parameter evaluated directly upstream ($V_{in,e}$) and downstream ($V_{out,e}\simeq V_{Ae}$) of the EDR. The ``$0$'' subscript indicates an asymptotic quantity. The dashed line is the separatrix dividing the upstream magnetic field from the reconnected field lines and exhaust.}
\end{figure}

\begin{equation}\label{eqn:rx_rate}
    \mathcal{R}(\delta/\ell) \simeq \frac{\delta}{\ell} \bigg( \frac{1-(\delta/\ell)^2}{1+(\delta/\ell)^2}\bigg)^2 \sqrt{1-\bigg(\frac{\delta}{\ell}\bigg)^2}
\end{equation}

Note that this expression is maximized by $\delta/\ell$ and can be evaluated using the local aspect ratio measured with Equation (\ref{eqn:aspect_ratio}), which is an independent quantity that can take any value.

\section{Methodology}

The magnetic field gradient terms used to evaluate Equation (\ref{eqn:aspect_ratio}) are very well measured by the MMS mission, which consists of four spacecraft that simultaneously measure magnetic field vectors at 128 Hz and calibrated to within 0.1 nT \cite{mms_fields_torbert2016}. The time-varying $\nabla \mathbf{B}(t)$ matrix is estimated at each time using the linear gradient method, which assumes that the spatial gradients within the volume of the spacecraft tetrahedron are linear  \cite{Dunlop2008}. In general, the ratio of the intermediate to largest eigenvalues of the symmetric matrix $(\nabla \mathbf{B}(t))(\nabla \mathbf{B}(t))^T$ yields the square of the current sheet aspect ratio \cite{rezeau2018}. At the X-point for symmetric anti-parallel reconnection, the largest and intermediate eigenvalues correspond exactly to $\lambda_N \sim (\partial B_L/\partial N)^2$ and $\lambda_L \sim (\partial B_N/\partial L)^2$. The diffusion region aspect ratio can then be written $\delta/\ell \simeq (\lambda_N/\lambda_L)^{1/4}$ which we refer to as the \textit{eigenvalue method}. Uncertainties arising from transforming the $\nabla \mathbf{B}(t)$ matrix from geocentric-solar coordinate system to a current sheet normal LMN coordinate system are avoided by using the minimum directional derivative of \textbf{B} (MDD-B) method, which determines the gradient terms in a time-varying coordinate system as eigenvectors of the matrix $(\nabla \mathbf{B}(t))(\nabla 
\mathbf{B}(t))^T$ \cite{Shi_2005,Shi_2019}. Therefore, when time-averaging the quantity $\delta/\ell$ calculated using Equation (\ref{eqn:aspect_ratio}) and gradients from MMS, which we will call $\langle\delta/\ell\rangle$ then the standard deviation over the same interval $\sigma[\delta/\ell]$ will capture any temporal or spatial variation in the current sheet structure. We can evaluate Equation (\ref{eqn:rx_rate}) with the time averaged aspect ratio $\langle \delta/\ell \rangle$ indicated as $\mathcal{R}\langle\delta/\ell\rangle$ or time-average $\mathcal{R}(\delta/\ell)$, written as $\langle \mathcal{R}(\delta/\ell)\rangle$. In most cases, these two expressions are equivalent. However, since Equation (\ref{eqn:rx_rate}) isn't single-valued there can be slight differences between the two with $\mathcal{R}\langle \delta/\ell \rangle$ tending to be slightly larger than $\langle \mathcal{R}(\delta/\ell)\rangle$. In this Letter, we calculate the rate for all events as $\mathcal{R}\langle\delta/\ell\rangle$ unless otherwise noted.   

\begin{figure}
\includegraphics{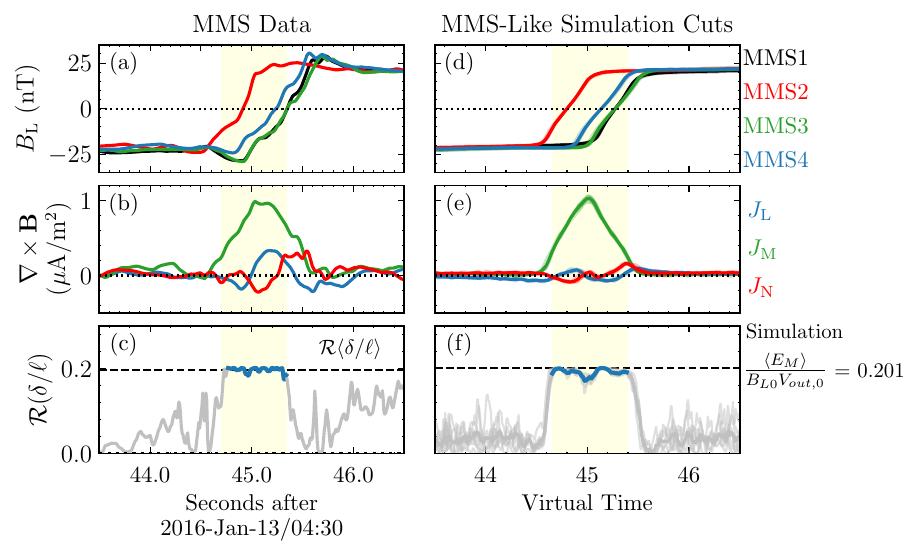}
\caption{\label{fig:2} (a) Reconnecting component $B_L$ for all 4 MMS spacecraft (b) the current density calculated using the multi-spacecraft linear gradient technique (c) the time-dependent normalized rate showing the spatially uniform region where $\delta/\ell$ takes the unique value within the current sheet that maximizes Equation (\ref{eqn:rx_rate}), with the horizontal line showing the average 0.197 (d-f) show the same quantities for one-dimensional cuts along MMS-like virtual spacecraft trajectories through a particle-in-cell simulation of the event. The highlighted region shows the $J_M > 0$ peak used for determining the time-averaged $\mathcal{R}\langle \delta/\ell\rangle$ and the associated uncertainty. The solid blue line in (f) shows the average of spacecraft cuts 12$d_e$ from the X-line between $t\Omega_{ci}=50$ and $t\Omega_{ci}=90$ (gray lines), with the horizontal line at 0.201 showing simulation normalized reconnection electric field.}
\end{figure}

\section{Method Validation for Guide Field and Asymmetric Reconnection}

The eigenvalue method has previously been demonstrated to accurately recover the EDR aspect ratio and normalized reconnection rate for symmetric anti-parallel reconnection \cite{Heuer_2022}. Since we are performing a multi-event study with events occurring across a wide range of reconnection regimes it is necessary to validate that the eigenvalue method is sufficiently general for reconnection with added guide field and asymmetries. For the strong guide field case, Figure 2(a-c) shows MMS data for 13 January 2016, a symmetric magnetosheath event with a guide field of 2.21, the largest in our dataset. In Figure 2(d-f), the MMS data is compared with cuts through a particle-in-cell (PIC) simulation of the event. One-dimensional cuts of the full magnetic field vector along four MMS-like virtual spacecraft trajectories (Figure 2d shows the four virtual $B_L$ cuts) were used to reconstruct a simulated $\nabla \mathbf{B}(t)$ by applying the same multi-spacecraft linear gradient method as with the real MMS data. We find excellent agreement between the time-dependent $\mathcal{R}\langle \delta/\ell \rangle$ from real MMS data (Figure 2c) and the normalized rate from evaluating Equation (2) with an aspect ratio calculated using with gradient terms from the simulated $\nabla \mathbf{B}(t)$ (Figure 2f). 

Within the current sheet ($J_M > 0$ in Figures 2b and 2c), the local time-dependent aspect ratio from Equation (1) is steady at the unique value which maximizes $\mathcal{R}$ when calculated with $\nabla \mathbf{B}(t)$ terms from both the MMS data (Figure 2c) and the simulated one-dimensional magnetic field cuts (Figure 2f). This produces a spatially and temporally uniform region within the current sheet (highlighted region in Figure 2) which can be time-averaged to obtain a single-value normalized reconnection rate $\mathcal{R}\langle \delta/\ell \rangle$ for the event. Since $\delta/\ell$ is time-dependent this captures any spatial or temporal uncertainty. The blue line in Figure 2(f) is the average of $\mathcal{R}\langle \delta/\ell \rangle$ calculated using magnetic fields cuts along virtual spacecraft trajectories with the tetrahedron barycenter crossing $12 d_e$ from the X-line between $t\Omega_{ci}=50$ and $t\Omega_{ci}=90$, which are shown in light blue. Within the current sheet over this time period there is almost no spatial or temporal variability. When compared with the MMS data this event is very steady, although this is not necessarily common for magnetosheath reconnection.

We find the single-value normalized rate measured by MMS for the 13 January 2016 event shown in Figure 2 is $\mathcal{R}\langle \delta/\ell \rangle=0.197 \pm0.006$ is within 2\% of the simulation normalized reconnection electric  $\langle E_{M}\rangle / (B_{L0} V_{out,0}) = 0.201$. $\langle E_{M}\rangle$ is the spatially averaged reconnection electric field, and $B_{L0}$ is the asymptotic reconnecting component of the magnetic field. The asymptotic ion Alfven speed $V_{Ai0}$ was not used because a magnetic island prevented one outflow jet to fully develop. Therefore, the ratio of the average outflow velocity for both jets and $V_{Ai0}$ was used instead, a technique that has been applied in previous studies for calculating the normalized rate in PIC simulations where one or both outflow jets are underdeveloped \cite{TKM_Nakamura2018}. The EDR aspect ratio was also determined spatially from $E_{\parallel} > 0$ near the X-line and used to evaluate Equation (\ref{eqn:rx_rate}), which resulted in a normalized rate of 0.198. The strong agreement between these independently evaluated normalized rates shows that the eigenvalue method is valid in the guide field regime. The eigenvalue method is validated for asymmetric reconnection in the same way as the symmetric guide field case. We analyzed a PIC simulation with magnetic field asymmetry $B_{L1}/B_{L2} = 0.3$ and density asymmetry $n_1/n_2 = 3$ \cite{liu2018,Genestreti2020} and found the difference between the simulation normalized reconnection electric field and normalized rate $\mathcal{R}(\delta/\ell)$ at the X-point to be $<10\%$. This demonstrates that the eigenvalue method is valid for asymmetric reconnection, and that we can still accurately recover the normalized rate with Equation (2) for the range of density and magnetic field asymmetry in our dataset. (See Appendix for initial conditions for both simulations). In addition to simulation analysis, the eigenvalue method was validated for symmetric anti-parallel reconnection in previous studies with comparisons of $\mathcal{R}\langle \delta/\ell \rangle$ measured by MMS to previously published values of the normalized reconnection electric field \cite{Heuer_2022}. However, for the asymmetric and strong guide field regimes direct comparison of $\mathcal{R}\langle \delta/\ell \rangle$ to the normalized reconnection electric field is not as straightforward since the electric fields at the magnetopause and magnetosheath tend to be smaller and closer to measurement uncertainties \cite{Pritchard2023}. One previous study used advantageous MMS trajectories through an EDR on 15 April 2018 to estimate the normalized rate as $V_{in,e}/V_{Ae} = 0.120 \pm 20\%$ \cite{burch2020_rxRate}, which is within uncertainties of our value $0.132 \pm 0.038$.

\begin{figure}
\includegraphics{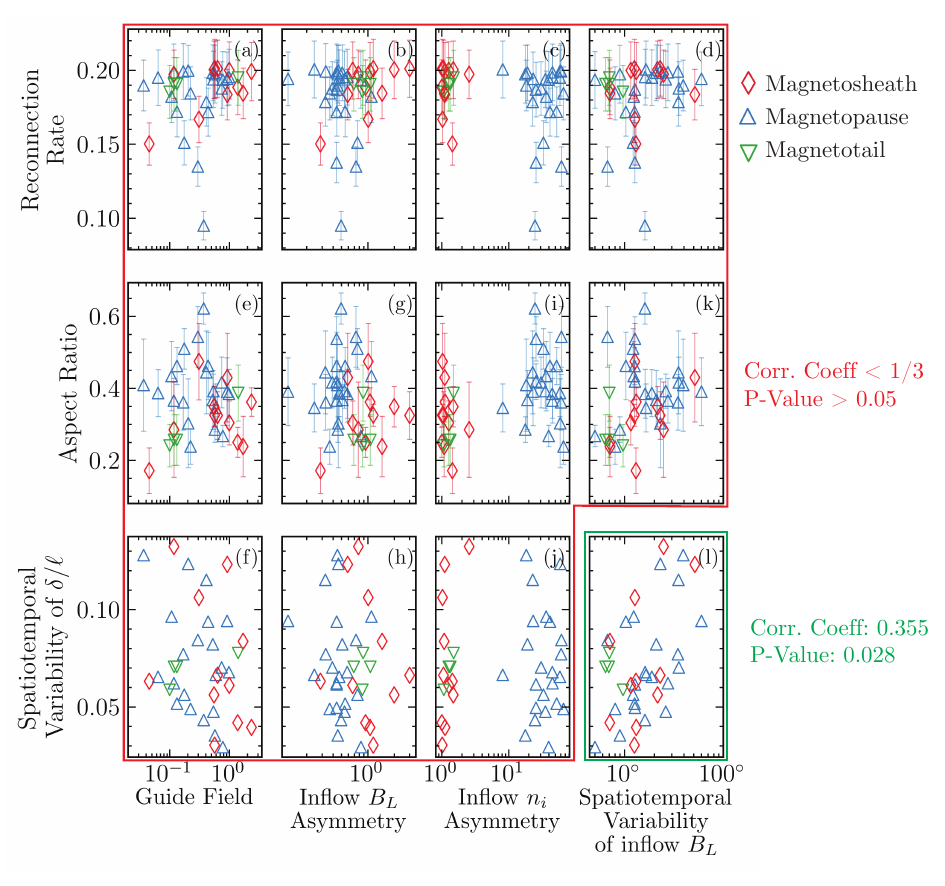}
\caption{\label{fig:3} Multi-event analysis showing the time-averaged normalized reconnection rate $\mathcal{R}\langle \delta/\ell \rangle$, the aspect ratio $\delta/\ell$, and the spatiotemporal variability of $\delta/\ell$ within the current sheet as a function of (a,e,f) normalized guide field $B_G=B_{M0}/B_{L0}$ (b,g,h) inflow asymmetry in the reconnecting component of the magnetic field $B_{L1}/B_{L2}$ (c,i,j) inflow ion density asymmetry $n_{1}/n_{2}$ and (d,k,l) spatiotemporal variability of the inflow reconnecting component of the magnetic field $\delta \theta$.}
\end{figure}

\section{Multi-Event Study}

The 38 events for the multi-event study were selected from previously identified EDR or near-EDR encounters observed by MMS, spanning the range of typical magnetospheric conditions. The largest subset of events is 23 asymmetric magnetopause reconnection events with varying guide fields \cite{Webster2018,Genestreti_2022,burch2020_rxRate,Pritchard2019}, followed by 11 magnetosheath reconnection events with varying guide fields \cite{wilder2018}, and symmetric magnetotail reconnection, with only 4 events \cite{torbert_science2018,farrugia2021,Zhou2019,hosner2024}. Events were discarded when no steady asymptotic inflow region could be identified, the local EDR parameters could not be clearly associated with the asymptotic inflow conditions, or if the MMS satellites were not in a tetrahedron formation. We note that the multiple studies combined in our event list do not all use identical criteria for EDR identification. However, some (but not all) examples include rapid magnetic field reversals, positive non-ideal energy conversion $\mathbf{J}\cdot\mathbf{E}^{\prime}$ \cite{Zenitani2011}, and large electron pressure non-gyrotropy \cite{Swisdak2016}. For all events we use the subscript ``1'' to indicate the high plasma density side of the reconnecting current sheet and the subscript ``2'' for the low plasma density side. Hence, the inflow plasma density asymmetry is defined $n_1/n_2$ and inflow asymmetry of the reconnecting component of the magnetic field as $B_{L1}/B_{L2}$. While for magnetopause reconnection events we always have $n_1/n_2>1$ and $B_{L1}/B_{L2}<1$ this is not the case for magnetosheath and magnetotail reconnection. For consistency, we use the same notation for all events.

The single-value normalized reconnection rate $\mathcal{R}\langle \delta/\ell \rangle$ for each event was obtained by evaluating Equation (\ref{eqn:rx_rate}) with the time-dependent aspect ratio averaged over the current sheet interval, estimated using the out-of-plane current $J_M$, as shown in Figure 2. The associated uncertainty in rate is taken to be 10\% of $\mathcal{R}\langle \delta/\ell \rangle$, the upper-bound determined from simulation analysis. The spatiotemporal variability in aspect ratio is found using the standard deviation of $\delta/\ell$ over the current sheet interval. The normalized guide magnetic field strength is defined $B_G = B_{M0}/B_{L0}$ where $B_{L0}$ is the hybrid reconnecting magnetic field component $B_{L0} = 2(B_{L1} B_{L2})/(B_{L1}+B_{L2})$ using the local LMN coordinates system shown in Figure (1)\cite{CassakShay2007}. The hybrid time-averaged guide field $B_{M0}$ follows the same form \cite{Genestreti_2022}. The temporal variation of the upstream magnetic field is defined $\delta \theta = \langle \mathrm{acos}(\hat{b}\cdot \langle \hat{b}\rangle)\rangle$ \cite{Genestreti_2022}. For the magnetopause reconnection events, the magnetosheath side was used because field variability in magnetospheric fields is low (below $10^{\circ}$ for all events). For magnetotail and magnetosheath events, an average of $\delta \theta$ on high and low-density sides was used. We test the strength of the correlation between the different parameters using Spearman's $\rho$ coefficient, which was chosen because (a) it is less sensitive to outliers, and (b) we have no functional form for any underlying relationship, and the Spearman correlation coefficient only assumes a monotonic relationship. We classify a correlation with $|\rho|<1/3$ as weak, $1/3<|\rho|<2/3$ as moderate and $|\rho|>2/3$ as strong. To estimate the significance of a correlation, we assume a null hypothesis of no correlation ($\rho$ = 0) and set a P-value threshold of 5\% as significant evidence for rejecting the null hypothesis in favor of a correlation.       

\section{Results of Multi-Event Analysis and Discussion}

The normalized reconnection rates, aspect ratio, and aspect ratio variability for all events are shown in Figure 3 as a function of guide field (Figure 3a, 3e, 3f), inflow $B_L$ asymmetry (Figure 3b, 3g, 3h), inflow density asymmetry (Figure 3c, 3i, 3j) and spatiotemporal angular variability of inflow $B_L$ (Figure 3d, 3k, 3l). Overall, we find minimal variation in the normalized reconnection rates between events, with $\mathcal{R}\langle \delta/\ell \rangle$ almost exclusively falling between 0.1 and 0.2 (with one outlier). The average uncertainty in normalized reconnection rate for the events shown in Figure 3 is below 12\%. Without controlling for any additional variables, we find no clear evidence for any statistically significant correlation between the normalized reconnection rate and guide field, inflow ion density asymmetry, inflow $B_L$ asymmetry, or the inflow $B_L$ angular variability (Figure 3b, 3c, and 3d). We also find no statistically significant correlation between reconnection rate or aspect ratio $\delta/\ell$ with any of the upstream parameters (3c, 3g, 3i, 3k), or the spatiotemporal variability of $\delta/\ell$ on guide field, $B_L$ asymmetry, or density asymmetry. We do find a moderate positive correlation between the spatiotemporal variability of the aspect ratio $\delta/\ell$ and the spatiotemporal angular variability of the inflow $B_L$ with a correlation coefficient of 0.355 and a P-value of 0.028 (Figure 3l). This would suggest that while the unnormalized rate of reconnection may change, the normalized reconnection rate is constant throughout the magnetosphere across the typical ranges of density asymmetry, $B_L$ asymmetry, and angular variability in $B_L$ but that the current sheet stability is affected by angular variability in the upstream magnetic field. When the correlation analysis is performed using $\langle\mathcal{R}(\delta/\ell)\rangle$ instead of $\mathcal{R}\langle\delta/\ell\rangle$ we find a weak but statistically significant correlation between reconnection rate and guide field. Since the significance of this correlation when using $\mathcal{R}\langle\delta/\ell\rangle$ is well below our threshold for acceptance, we consider this apparent correlation as a statistical anomaly since the difference in reconnection rate calculations is not substantial. 

Interestingly, there is substantially more variability in the reconnection rate among events with the smaller ($B_G<0.5$) versus larger ($B_G>0.5$) guide fields, with the standard deviation in reconnection rate being an order of magnitude larger in the former subset than the latter, with an average uncertainty twice as large. The average normalized rate for events with $B_G > 0.5$ is 15\% higher than those below. This difference suggests that there is an additional uncontrolled variable arising at lower guide fields which is affecting either the normalized reconnection rate, or the stability and structure of the current sheet, and that $B_G \simeq 0.5$ is the approximate critical value. There is a moderate negative correlation between guide field and spatiotemporal aspect ratio variability supporting stabilization of the current sheet for higher guide fields, but is not statistically significant. It is worth noting that $B_G \simeq 0.5$ is the point where electron dissipation switches from perpendicular to parallel-dominated, as found in a study of the same magnetosheath events considered here and in parametric simulation studies \cite{wilder2018,Pucci2018}. Simulation studies have found that the normalized reconnection rate should be independent of electron dissipation mechanism, since they carry a small fraction of the total energy in the system \cite{shay1998}. However, the transition from perpendicular to parallel electron dissipation may introduce some current sheet instability which could cause time dependent fluctuations on top of a constant rate. 
One caveat on the guide field correlation is that the highest guide field events all occur within the magnetosheath, which is expected due to the low guide fields in the magnetotail and the diamagnetic suppression of low magnetic shear reconnection at the magnetopause. However, reconnection in the turbulent magnetosheath is often strongly driven and the regime most likely to violate our steady-state assumption. During the small-scale impulsive reconnection typical of the magnetosheath, the larger-scale fluid stresses haven’t had time to couple with the kinetic scales and so the normalized reconnection rate may go as the EDR aspect ratio. The rate calculated using Equation (2) has a maximum value of 0.2 but simulation and laboratory studies of the transition between electron-only and ion-coupled reconnection have found that the normalized rate increases as the spatial scale of reconnection drops, with predictions of the rate for electron-only reconnection being between 0.4 and 0.5 \cite{Greess2022,Liu2025}. No known electron-only events were included in this study and further work is needed to investigate this possibility. 
The spatiotemporal aspect ratio variability correlation suggests that either spatial or temporal variation in the direction of the inflow $B_L$ affects the stability of the current sheet on the time scale of MMS observations. Previous studies have found that three-dimensional descriptions of the EDR can be needed even for magnetotail reconnection, which could generate magnetic field perturbations that would be captured in the aspect ratio variability. However, determining whether there is a correlation between $\delta\theta$ and the current sheet dimensionality is beyond the scope of this Letter. We also note that the event with the lowest reconnection rate has previously been identified as a recently formed secondary X-line \cite{Hasegawa2023_secondaryXline}. Equation (2) was derived for the steady-state case, and if outliers have a time dependence with significant deviations from force balance, then the normalized reconnection rate from averaging within the current sheet may be inaccurate. While this explanation would give the required variability for single events, spacecraft observations at the dayside magnetopause have found that distribution and frequency of magnetic islands and secondary X-lines does not depend on the guide field \cite{vines2017}. 

\section{Conclusions and Future Work}
In this Letter we have performed a multi-event study of the normalized reconnection rate controlling for guide field and density and magnetic field asymmetry across the three regimes of reconnection observed in Earth's magnetosphere. We also test for correlations with the current sheet aspect ratio and the spatiotemporal variability of the current sheet aspect ratio. We find a moderate but statistically significant positive correlations between the spatiotemporal variability in the aspect ratio and the spatiotemporal angular variability in the upstream reconnecting component of the magnetic field, but the normalized reconnection rate appears independent of the upstream plasma conditions. In the future, additional events could be included to expand and extend the parameter space and allow controlling for additional variables, and improvements could be made to the reconnection rate model used to better capture time dependence and electron-only reconnection. This is the first study to use the novel and accurate magnetic field gradient method to evaluate the dependence of the normalized reconnection rate over the three dominant regimes of reconnection within the magnetosphere with in situ measurements. These results represent an improvement in the state-of-the-art and may provide insight into the rate at which energy from the solar wind enters the Earth's magnetosphere, which may be significant in predicting the terrestrial effects of space weather events. 

\appendix

\section{Simulation Setup}

This study uses two 2.5D-dimensional VPIC code \cite{Bowers2009}. The first simulation was performed for this event with initial conditions chosen to match the 13 January 2016 event analyzed for our guide field case. The simulation guide field was chosen to be $B_g=2.1$ with a mass ratio $m_i/m_e=100$ and uniform temperature with temperature ratio $T_i/T_e =8$. The plasma density and reconnecting component of the magnetic field are symmetric across the current sheet. The ratio of electron plasma to gyro frequencies is $\omega_{pe}/\Omega_{ci}=2$. Virtual spacecraft trajectories were selected to match by best fit of $B_L$ with relative positions chosen from 3D magnetic field reconstructions. Simulation output is available from the authors upon reasonable request. 

The second simulation was performed for a previous study \cite{Genestreti2020}. The mass ratio is $m_i/m_e=75$. The current sheet an an initial profile of $B_L(N) = B_1(0.5-S(N))$ where $B_1=-B_M$ and $S(N)=\tanh(N/\delta)$. The initial current sheet half thickness is $\delta =0.5 d_i$ and density profile $n(N)=n_1[1-(S^2-S)/3]$ with asymmetry $n_{sh}/n_{sp}=3$ (where $n_{sh}$ corresponds to $n_1$ in our event analysis). The temperature is uniform with ratio $T_i/T_e = 5$. 

%



%
%

\section*{Open Research Section}The MMS FIELDS and FPI data used in this study are publicly available from the Science Data Center (SDC) at https://lasp.colorado.edu/mms/sdc/public/about/browse. Routines from the pySPEDAS and pyTplot libraries used to process the MMS data analyzed in this paper \cite{grimes2022} are freely available at https://github.com/spedas/pyspedas. 

The asymmetric simulation outputs performed for a previous study \cite{Genestreti2020} are available as ``Run 2'' on Zenodo.org at https://doi.org/10.5281/zenodo.3945043. The simulation of the symmetric guide field event performed for this study is also available on Zenodo.org at https://zenodo.org/uploads/15009472.

\acknowledgments
We would like to thank J.M.H Beedle, K.R Pritchard, and A. Marshall for valuable discussion. We also thank all MMS SITL volunteers and the many authors who found the reconnection events used in this study. S.V.H., K.J.G., and R.B.T. were supported by NASA's MMS FIELDS Grant NNG04EB99C. J.R.S was supported by NASA Grants 80NSSC23K1152 and 80NSSC23K1600.

%
%

\bibliography{Scaling_of_The_Normalized_Magnetic_Reconnection_Rate.bib}

\begin{thebibliography}{}

\bibitem [\protect \citeauthoryear {%
Baker%
, Pulkkinen%
, Angelopoulos%
, Baumjohann%
\BCBL {}\ \BBA {} McPherron%
}{%
Baker%
\ \protect \BOthers {.}}{%
{\protect \APACyear {1996}}%
}]{%
Baker1996}
\APACinsertmetastar {%
Baker1996}%
\begin{APACrefauthors}%
Baker, D\BPBI N.%
, Pulkkinen, T\BPBI I.%
, Angelopoulos, V.%
, Baumjohann, W.%
\BCBL {}\ \BBA {} McPherron, R\BPBI L.%
\end{APACrefauthors}%
\unskip\
\newblock
\APACrefYearMonthDay{1996}{}{}.
\newblock
{\BBOQ}\APACrefatitle {Neutral line model of substorms: Past results and present view} {Neutral line model of substorms: Past results and present view}.{\BBCQ}
\newblock
\APACjournalVolNumPages{J. Geophys. Res. Space Phys.}{101}{A6}{12975-13010}.
\newblock
\begin{APACrefDOI} \doi{https://doi.org/10.1029/95JA03753} \end{APACrefDOI}
\PrintBackRefs{\CurrentBib}

\bibitem [\protect \citeauthoryear {%
{Bowers}%
, {Albright}%
, {Yin}%
, {Bergen}%
\BCBL {}\ \BBA {} {Kwan}%
}{%
{Bowers}%
\ \protect \BOthers {.}}{%
{\protect \APACyear {2008}}%
}]{%
Bowers2009}
\APACinsertmetastar {%
Bowers2009}%
\begin{APACrefauthors}%
{Bowers}, K\BPBI J.%
, {Albright}, B\BPBI J.%
, {Yin}, L.%
, {Bergen}, B.%
\BCBL {}\ \BBA {} {Kwan}, T\BPBI J\BPBI T.%
\end{APACrefauthors}%
\unskip\
\newblock
\APACrefYearMonthDay{2008}{{\APACmonth{05}}}{}.
\newblock
{\BBOQ}\APACrefatitle {{Ultrahigh performance three-dimensional electromagnetic relativistic kinetic plasma simulationa)}} {{Ultrahigh performance three-dimensional electromagnetic relativistic kinetic plasma simulationa)}}.{\BBCQ}
\newblock
\APACjournalVolNumPages{Phys. Plasmas}{15}{5}{055703}.
\newblock
\begin{APACrefDOI} \doi{10.1063/1.2840133} \end{APACrefDOI}
\PrintBackRefs{\CurrentBib}

\bibitem [\protect \citeauthoryear {%
{Burch}%
\ \protect \BOthers {.}}{%
{Burch}%
\ \protect \BOthers {.}}{%
{\protect \APACyear {2022}}%
}]{%
burch2022_edr_inflow_tail}
\APACinsertmetastar {%
burch2022_edr_inflow_tail}%
\begin{APACrefauthors}%
{Burch}, J\BPBI L.%
, {Hesse}, M.%
, {Webster}, J\BPBI M.%
, {Genestreti}, K\BPBI J.%
, {Torbert}, R\BPBI B.%
, {Denton}, R\BPBI E.%
\BDBL {}{Pollock}, C\BPBI J.%
\end{APACrefauthors}%
\unskip\
\newblock
\APACrefYearMonthDay{2022}{}{}.
\newblock
{\BBOQ}\APACrefatitle {{The EDR inflow region of a reconnecting current sheet in the geomagnetic tail}} {{The EDR inflow region of a reconnecting current sheet in the geomagnetic tail}}.{\BBCQ}
\newblock
\APACjournalVolNumPages{Phys. Plasmas}{29}{5}{052903}.
\newblock
\begin{APACrefDOI} \doi{10.1063/5.0083169} \end{APACrefDOI}
\PrintBackRefs{\CurrentBib}

\bibitem [\protect \citeauthoryear {%
{Burch}%
, {Moore}%
, {Torbert}%
\BCBL {}\ \BBA {} {Giles}%
}{%
{Burch}%
\ \protect \BOthers {.}}{%
{\protect \APACyear {2016}}%
}]{%
mms_burch2016}
\APACinsertmetastar {%
mms_burch2016}%
\begin{APACrefauthors}%
{Burch}, J\BPBI L.%
, {Moore}, T\BPBI E.%
, {Torbert}, R\BPBI B.%
\BCBL {}\ \BBA {} {Giles}, B\BPBI L.%
\end{APACrefauthors}%
\unskip\
\newblock
\APACrefYearMonthDay{2016}{}{}.
\newblock
{\BBOQ}\APACrefatitle {{Magnetospheric Multiscale Overview and Science Objectives}} {{Magnetospheric Multiscale Overview and Science Objectives}}.{\BBCQ}
\newblock
\APACjournalVolNumPages{Space Sci. Rev.}{199}{1-4}{5-21}.
\newblock
\begin{APACrefDOI} \doi{10.1007/s11214-015-0164-9} \end{APACrefDOI}
\PrintBackRefs{\CurrentBib}

\bibitem [\protect \citeauthoryear {%
Burch%
\ \protect \BOthers {.}}{%
Burch%
\ \protect \BOthers {.}}{%
{\protect \APACyear {2020}}%
}]{%
burch2020_rxRate}
\APACinsertmetastar {%
burch2020_rxRate}%
\begin{APACrefauthors}%
Burch, J\BPBI L.%
, Webster, J\BPBI M.%
, Hesse, M.%
, Genestreti, K\BPBI J.%
, Denton, R\BPBI E.%
, Phan, T\BPBI D.%
\BDBL {}Paschmann, G.%
\end{APACrefauthors}%
\unskip\
\newblock
\APACrefYearMonthDay{2020}{}{}.
\newblock
{\BBOQ}\APACrefatitle {Electron Inflow Velocities and Reconnection Rates at Earth's Magnetopause and Magnetosheath} {Electron inflow velocities and reconnection rates at earth's magnetopause and magnetosheath}.{\BBCQ}
\newblock
\APACjournalVolNumPages{Geophysical. Res. Lett.}{47}{17}{e2020GL089082}.
\newblock
\begin{APACrefDOI} \doi{https://doi.org/10.1029/2020GL089082} \end{APACrefDOI}
\PrintBackRefs{\CurrentBib}

\bibitem [\protect \citeauthoryear {%
Cassak%
, Liu%
\BCBL {}\ \BBA {} Shay%
}{%
Cassak%
\ \protect \BOthers {.}}{%
{\protect \APACyear {2017}}%
}]{%
Cassak_Liu_Shay_2017}
\APACinsertmetastar {%
Cassak_Liu_Shay_2017}%
\begin{APACrefauthors}%
Cassak, P\BPBI A.%
, Liu, Y\BHBI H.%
\BCBL {}\ \BBA {} Shay, M.%
\end{APACrefauthors}%
\unskip\
\newblock
\APACrefYearMonthDay{2017}{}{}.
\newblock
{\BBOQ}\APACrefatitle {A review of the 0.1 reconnection rate problem} {A review of the 0.1 reconnection rate problem}.{\BBCQ}
\newblock
\APACjournalVolNumPages{J. Plasma Phys.}{83}{5}{715830501}.
\newblock
\begin{APACrefDOI} \doi{10.1017/S0022377817000666} \end{APACrefDOI}
\PrintBackRefs{\CurrentBib}

\bibitem [\protect \citeauthoryear {%
{Cassak}%
\ \BBA {} {Shay}%
}{%
{Cassak}%
\ \BBA {} {Shay}%
}{%
{\protect \APACyear {2007}}%
}]{%
CassakShay2007}
\APACinsertmetastar {%
CassakShay2007}%
\begin{APACrefauthors}%
{Cassak}, P\BPBI A.%
\BCBT {}\ \BBA {} {Shay}, M\BPBI A.%
\end{APACrefauthors}%
\unskip\
\newblock
\APACrefYearMonthDay{2007}{{\APACmonth{10}}}{}.
\newblock
{\BBOQ}\APACrefatitle {{Scaling of asymmetric magnetic reconnection: General theory and collisional simulations}} {{Scaling of asymmetric magnetic reconnection: General theory and collisional simulations}}.{\BBCQ}
\newblock
\APACjournalVolNumPages{Phys. Plasmas}{14}{10}{102114}.
\newblock
\begin{APACrefDOI} \doi{10.1063/1.2795630} \end{APACrefDOI}
\PrintBackRefs{\CurrentBib}

\bibitem [\protect \citeauthoryear {%
{Chen}%
\ \protect \BOthers {.}}{%
{Chen}%
\ \protect \BOthers {.}}{%
{\protect \APACyear {2017}}%
}]{%
LJ_Chen_2017}
\APACinsertmetastar {%
LJ_Chen_2017}%
\begin{APACrefauthors}%
{Chen}, L\BPBI J.%
, {Hesse}, M.%
, {Wang}, S.%
, {Gershman}, D.%
, {Ergun}, R\BPBI E.%
, {Burch}, J.%
\BDBL {}{Avanov}, L.%
\end{APACrefauthors}%
\unskip\
\newblock
\APACrefYearMonthDay{2017}{{\APACmonth{05}}}{}.
\newblock
{\BBOQ}\APACrefatitle {{Electron diffusion region during magnetopause reconnection with an intermediate guide field: Magnetospheric multiscale observations}} {{Electron diffusion region during magnetopause reconnection with an intermediate guide field: Magnetospheric multiscale observations}}.{\BBCQ}
\newblock
\APACjournalVolNumPages{J. Geophys. Res. Space Phys}{122}{5}{5235-5246}.
\newblock
\begin{APACrefDOI} \doi{10.1002/2017JA024004} \end{APACrefDOI}
\PrintBackRefs{\CurrentBib}

\bibitem [\protect \citeauthoryear {%
{Comisso}%
\ \BBA {} {Bhattacharjee}%
}{%
{Comisso}%
\ \BBA {} {Bhattacharjee}%
}{%
{\protect \APACyear {2016}}%
}]{%
Comisso_Bhattacharjee_2016}
\APACinsertmetastar {%
Comisso_Bhattacharjee_2016}%
\begin{APACrefauthors}%
{Comisso}, L.%
\BCBT {}\ \BBA {} {Bhattacharjee}, A.%
\end{APACrefauthors}%
\unskip\
\newblock
\APACrefYearMonthDay{2016}{{\APACmonth{12}}}{}.
\newblock
{\BBOQ}\APACrefatitle {{On the value of the reconnection rate}} {{On the value of the reconnection rate}}.{\BBCQ}
\newblock
\APACjournalVolNumPages{J. Plasma Phys.}{82}{6}{595820601}.
\newblock
\begin{APACrefDOI} \doi{10.1017/S002237781600101X} \end{APACrefDOI}
\PrintBackRefs{\CurrentBib}

\bibitem [\protect \citeauthoryear {%
{Dunlop}%
\ \BBA {} {Eastwood}%
}{%
{Dunlop}%
\ \BBA {} {Eastwood}%
}{%
{\protect \APACyear {2008}}%
}]{%
Dunlop2008}
\APACinsertmetastar {%
Dunlop2008}%
\begin{APACrefauthors}%
{Dunlop}, M\BPBI W.%
\BCBT {}\ \BBA {} {Eastwood}, J\BPBI P.%
\end{APACrefauthors}%
\unskip\
\newblock
\APACrefYearMonthDay{2008}{{\APACmonth{01}}}{}.
\newblock
{\BBOQ}\APACrefatitle {{The Curlometer and Other Gradient Based Methods}} {{The Curlometer and Other Gradient Based Methods}}.{\BBCQ}
\newblock
\APACjournalVolNumPages{ISSI Scientific Reports Series}{8}{}{17-26}.
\PrintBackRefs{\CurrentBib}

\bibitem [\protect \citeauthoryear {%
Farrugia%
\ \protect \BOthers {.}}{%
Farrugia%
\ \protect \BOthers {.}}{%
{\protect \APACyear {2021}}%
}]{%
farrugia2021}
\APACinsertmetastar {%
farrugia2021}%
\begin{APACrefauthors}%
Farrugia, C\BPBI J.%
, Rogers, A\BPBI J.%
, Torbert, R\BPBI B.%
, Genestreti, K\BPBI J.%
, Nakamura, T\BPBI K\BPBI M.%
, Lavraud, B.%
\BDBL {}Dors, I.%
\end{APACrefauthors}%
\unskip\
\newblock
\APACrefYearMonthDay{2021}{}{}.
\newblock
{\BBOQ}\APACrefatitle {An Encounter With the Ion and Electron Diffusion Regions at a Flapping and Twisted Tail Current Sheet} {An encounter with the ion and electron diffusion regions at a flapping and twisted tail current sheet}.{\BBCQ}
\newblock
\APACjournalVolNumPages{J. Geophys. Res. Space Phys.}{126}{3}{e2020JA028903}.
\newblock
\APACrefnote{e2020JA028903 2020JA028903}
\newblock
\begin{APACrefDOI} \doi{https://doi.org/10.1029/2020JA028903} \end{APACrefDOI}
\PrintBackRefs{\CurrentBib}

\bibitem [\protect \citeauthoryear {%
{Fuselier}%
\ \BBA {} {Lewis}%
}{%
{Fuselier}%
\ \BBA {} {Lewis}%
}{%
{\protect \APACyear {2011}}%
}]{%
FuselierAndLewis2011}
\APACinsertmetastar {%
FuselierAndLewis2011}%
\begin{APACrefauthors}%
{Fuselier}, S\BPBI A.%
\BCBT {}\ \BBA {} {Lewis}, W\BPBI S.%
\end{APACrefauthors}%
\unskip\
\newblock
\APACrefYearMonthDay{2011}{}{}.
\newblock
{\BBOQ}\APACrefatitle {{Properties of Near-Earth Magnetic Reconnection from In-Situ Observations}} {{Properties of Near-Earth Magnetic Reconnection from In-Situ Observations}}.{\BBCQ}
\newblock
\APACjournalVolNumPages{Space Sci. Rev.}{160}{1-4}{95-121}.
\newblock
\begin{APACrefDOI} \doi{10.1007/s11214-011-9820-x} \end{APACrefDOI}
\PrintBackRefs{\CurrentBib}

\bibitem [\protect \citeauthoryear {%
{Genestreti}%
\ \protect \BOthers {.}}{%
{Genestreti}%
\ \protect \BOthers {.}}{%
{\protect \APACyear {2022}}%
}]{%
Genestreti_2022}
\APACinsertmetastar {%
Genestreti_2022}%
\begin{APACrefauthors}%
{Genestreti}, K\BPBI J.%
, {Li}, X.%
, {Liu}, Y\BHBI H.%
, {Burch}, J\BPBI L.%
, {Torbert}, R\BPBI B.%
, {Fuselier}, S\BPBI A.%
\BDBL {}{Strangeway}, R\BPBI J.%
\end{APACrefauthors}%
\unskip\
\newblock
\APACrefYearMonthDay{2022}{{\APACmonth{08}}}{}.
\newblock
{\BBOQ}\APACrefatitle {{On the origin of ``patchy'' energy conversion in electron diffusion regions}} {{On the origin of ``patchy'' energy conversion in electron diffusion regions}}.{\BBCQ}
\newblock
\APACjournalVolNumPages{Phys. Plasmas}{29}{8}{082107}.
\newblock
\begin{APACrefDOI} \doi{10.1063/5.0090275} \end{APACrefDOI}
\PrintBackRefs{\CurrentBib}

\bibitem [\protect \citeauthoryear {%
Genestreti%
\ \protect \BOthers {.}}{%
Genestreti%
\ \protect \BOthers {.}}{%
{\protect \APACyear {2020}}%
}]{%
Genestreti2020}
\APACinsertmetastar {%
Genestreti2020}%
\begin{APACrefauthors}%
Genestreti, K\BPBI J.%
, Liu, Y\BHBI H.%
, Phan, T\BHBI D.%
, Denton, R\BPBI E.%
, Torbert, R\BPBI B.%
, Burch, J\BPBI L.%
\BDBL {}Eriksson, S.%
\end{APACrefauthors}%
\unskip\
\newblock
\APACrefYearMonthDay{2020}{}{}.
\newblock
{\BBOQ}\APACrefatitle {Multiscale Coupling During Magnetopause Reconnection: Interface Between the Electron and Ion Diffusion Regions} {Multiscale coupling during magnetopause reconnection: Interface between the electron and ion diffusion regions}.{\BBCQ}
\newblock
\APACjournalVolNumPages{J. Geophys. Res. Space Phys.}{125}{10}{e2020JA027985}.
\newblock
\APACrefnote{e2020JA027985 10.1029/2020JA027985}
\newblock
\begin{APACrefDOI} \doi{https://doi.org/10.1029/2020JA027985} \end{APACrefDOI}
\PrintBackRefs{\CurrentBib}

\bibitem [\protect \citeauthoryear {%
Genestreti%
\ \protect \BOthers {.}}{%
Genestreti%
\ \protect \BOthers {.}}{%
{\protect \APACyear {2018}}%
}]{%
genestreti2018}
\APACinsertmetastar {%
genestreti2018}%
\begin{APACrefauthors}%
Genestreti, K\BPBI J.%
, Nakamura, T\BPBI K\BPBI M.%
, Nakamura, R.%
, Denton, R\BPBI E.%
, Torbert, R\BPBI B.%
, Burch, J\BPBI L.%
\BDBL {}Russell, C\BPBI T.%
\end{APACrefauthors}%
\unskip\
\newblock
\APACrefYearMonthDay{2018}{}{}.
\newblock
{\BBOQ}\APACrefatitle {How Accurately Can We Measure the Reconnection Rate EM for the MMS Diffusion Region Event of 11 July 2017?} {How accurately can we measure the reconnection rate em for the mms diffusion region event of 11 july 2017?}{\BBCQ}
\newblock
\APACjournalVolNumPages{J. Geophys. Res. Space Phys.}{123}{11}{9130-9149}.
\newblock
\begin{APACrefDOI} \doi{https://doi.org/10.1029/2018JA025711} \end{APACrefDOI}
\PrintBackRefs{\CurrentBib}

\bibitem [\protect \citeauthoryear {%
Greess%
\ \protect \BOthers {.}}{%
Greess%
\ \protect \BOthers {.}}{%
{\protect \APACyear {2021}}%
}]{%
Greess2021}
\APACinsertmetastar {%
Greess2021}%
\begin{APACrefauthors}%
Greess, S.%
, Egedal, J.%
, Stanier, A.%
, Daughton, W.%
, Olson, J.%
, Lê, A.%
\BDBL {}Forest, C.%
\end{APACrefauthors}%
\unskip\
\newblock
\APACrefYearMonthDay{2021}{}{}.
\newblock
{\BBOQ}\APACrefatitle {Laboratory Verification of Electron-Scale Reconnection Regions Modulated by a Three-Dimensional Instability} {Laboratory verification of electron-scale reconnection regions modulated by a three-dimensional instability}.{\BBCQ}
\newblock
\APACjournalVolNumPages{J. Geophys. Res. Space Phys.}{126}{7}{e2021JA029316}.
\newblock
\begin{APACrefDOI} \doi{https://doi.org/10.1029/2021JA029316} \end{APACrefDOI}
\PrintBackRefs{\CurrentBib}

\bibitem [\protect \citeauthoryear {%
{Greess}%
\ \protect \BOthers {.}}{%
{Greess}%
\ \protect \BOthers {.}}{%
{\protect \APACyear {2022}}%
}]{%
Greess2022}
\APACinsertmetastar {%
Greess2022}%
\begin{APACrefauthors}%
{Greess}, S.%
, {Egedal}, J.%
, {Stanier}, A.%
, {Olson}, J.%
, {Daughton}, W.%
, {L{\^e}}, A.%
\BDBL {}{Forest}, C\BPBI B.%
\end{APACrefauthors}%
\unskip\
\newblock
\APACrefYearMonthDay{2022}{}{}.
\newblock
{\BBOQ}\APACrefatitle {{Kinetic simulations verifying reconnection rates measured in the laboratory, spanning the ion-coupled to near electron-only regimes}} {{Kinetic simulations verifying reconnection rates measured in the laboratory, spanning the ion-coupled to near electron-only regimes}}.{\BBCQ}
\newblock
\APACjournalVolNumPages{Phys. Plasmas}{29}{10}{102103}.
\newblock
\begin{APACrefDOI} \doi{10.1063/5.0101006} \end{APACrefDOI}
\PrintBackRefs{\CurrentBib}

\bibitem [\protect \citeauthoryear {%
Grimes%
\ \protect \BOthers {.}}{%
Grimes%
\ \protect \BOthers {.}}{%
{\protect \APACyear {2022}}%
}]{%
grimes2022}
\APACinsertmetastar {%
grimes2022}%
\begin{APACrefauthors}%
Grimes, E\BPBI W.%
, Harter, B.%
, Hatzigeorgiu, N.%
, Drozdov, A.%
, Lewis, J\BPBI W.%
, Angelopoulos, V.%
\BDBL {}Le~Contel, O.%
\end{APACrefauthors}%
\unskip\
\newblock
\APACrefYearMonthDay{2022}{}{}.
\newblock
{\BBOQ}\APACrefatitle {The Space Physics Environment Data Analysis System in Python} {The space physics environment data analysis system in python}.{\BBCQ}
\newblock
\APACjournalVolNumPages{Frontiers in Astronomy and Space Sciences}{Volume 9 - 2022}{}{}.
\newblock
\begin{APACrefURL} \url{https://www.frontiersin.org/journals/astronomy-and-space-sciences/articles/10.3389/fspas.2022.1020815} \end{APACrefURL}
\newblock
\begin{APACrefDOI} \doi{10.3389/fspas.2022.1020815} \end{APACrefDOI}
\PrintBackRefs{\CurrentBib}

\bibitem [\protect \citeauthoryear {%
{Hasegawa}%
\ \protect \BOthers {.}}{%
{Hasegawa}%
\ \protect \BOthers {.}}{%
{\protect \APACyear {2023}}%
}]{%
Hasegawa2023_secondaryXline}
\APACinsertmetastar {%
Hasegawa2023_secondaryXline}%
\begin{APACrefauthors}%
{Hasegawa}, H.%
, {Denton}, R\BPBI E.%
, {Dokgo}, K.%
, {Hwang}, K\BPBI J.%
, {Nakamura}, T\BPBI K\BPBI M.%
\BCBL {}\ \BBA {} {Burch}, J\BPBI L.%
\end{APACrefauthors}%
\unskip\
\newblock
\APACrefYearMonthDay{2023}{{\APACmonth{03}}}{}.
\newblock
{\BBOQ}\APACrefatitle {{Ion-Scale Magnetic Flux Rope Generated From Electron-Scale Magnetopause Current Sheet: Magnetospheric Multiscale Observations}} {{Ion-Scale Magnetic Flux Rope Generated From Electron-Scale Magnetopause Current Sheet: Magnetospheric Multiscale Observations}}.{\BBCQ}
\newblock
\APACjournalVolNumPages{J. Geophys. Res. Space Phys}{128}{3}{e2022JA031092}.
\newblock
\begin{APACrefDOI} \doi{10.1029/2022JA031092} \end{APACrefDOI}
\PrintBackRefs{\CurrentBib}

\bibitem [\protect \citeauthoryear {%
{Hesse}%
, {Neukirch}%
, {Schindler}%
, {Kuznetsova}%
\BCBL {}\ \BBA {} {Zenitani}%
}{%
{Hesse}%
\ \protect \BOthers {.}}{%
{\protect \APACyear {2011}}%
}]{%
Hesse2011}
\APACinsertmetastar {%
Hesse2011}%
\begin{APACrefauthors}%
{Hesse}, M.%
, {Neukirch}, T.%
, {Schindler}, K.%
, {Kuznetsova}, M.%
\BCBL {}\ \BBA {} {Zenitani}, S.%
\end{APACrefauthors}%
\unskip\
\newblock
\APACrefYearMonthDay{2011}{{\APACmonth{10}}}{}.
\newblock
{\BBOQ}\APACrefatitle {{The Diffusion Region in Collisionless Magnetic Reconnection}} {{The Diffusion Region in Collisionless Magnetic Reconnection}}.{\BBCQ}
\newblock
\APACjournalVolNumPages{Space Sci. Rev.}{160}{1-4}{3-23}.
\newblock
\begin{APACrefDOI} \doi{10.1007/s11214-010-9740-1} \end{APACrefDOI}
\PrintBackRefs{\CurrentBib}

\bibitem [\protect \citeauthoryear {%
{Heuer}%
\ \protect \BOthers {.}}{%
{Heuer}%
\ \protect \BOthers {.}}{%
{\protect \APACyear {2022}}%
}]{%
Heuer_2022}
\APACinsertmetastar {%
Heuer_2022}%
\begin{APACrefauthors}%
{Heuer}, S\BPBI V.%
, {Genestreti}, K\BPBI J.%
, {Nakamura}, T\BPBI K\BPBI M.%
, {Torbert}, R\BPBI B.%
, {Burch}, J\BPBI L.%
\BCBL {}\ \BBA {} {Nakamura}, R.%
\end{APACrefauthors}%
\unskip\
\newblock
\APACrefYearMonthDay{2022}{{\APACmonth{10}}}{}.
\newblock
{\BBOQ}\APACrefatitle {{Calculating the Electron Diffusion Region Aspect Ratio With Magnetic Field Gradients}} {{Calculating the Electron Diffusion Region Aspect Ratio With Magnetic Field Gradients}}.{\BBCQ}
\newblock
\APACjournalVolNumPages{Geophys. Res. Lett.}{49}{20}{}.
\newblock
\begin{APACrefDOI} \doi{10.1029/2022GL100652} \end{APACrefDOI}
\PrintBackRefs{\CurrentBib}

\bibitem [\protect \citeauthoryear {%
{Hosner}%
\ \protect \BOthers {.}}{%
{Hosner}%
\ \protect \BOthers {.}}{%
{\protect \APACyear {2024}}%
}]{%
hosner2024}
\APACinsertmetastar {%
hosner2024}%
\begin{APACrefauthors}%
{Hosner}, M.%
, {Nakamura}, R.%
, {Schmid}, D.%
, {Nakamura}, T\BPBI K\BPBI M.%
, {Panov}, E\BPBI V.%
, {Volwerk}, M.%
\BDBL {}{Fazakerley}, A\BPBI N.%
\end{APACrefauthors}%
\unskip\
\newblock
\APACrefYearMonthDay{2024}{}{}.
\newblock
{\BBOQ}\APACrefatitle {{Reconnection Inside a Dipolarization Front of a Diverging Earthward Fast Flow}} {{Reconnection Inside a Dipolarization Front of a Diverging Earthward Fast Flow}}.{\BBCQ}
\newblock
\APACjournalVolNumPages{J. Geophys. Res. Space Phys}{129}{1}{e2023JA031976}.
\newblock
\begin{APACrefDOI} \doi{10.1029/2023JA031976} \end{APACrefDOI}
\PrintBackRefs{\CurrentBib}

\bibitem [\protect \citeauthoryear {%
{Huba}%
}{%
{Huba}%
}{%
{\protect \APACyear {2005}}%
}]{%
Huba2005}
\APACinsertmetastar {%
Huba2005}%
\begin{APACrefauthors}%
{Huba}, J\BPBI D.%
\end{APACrefauthors}%
\unskip\
\newblock
\APACrefYearMonthDay{2005}{{\APACmonth{01}}}{}.
\newblock
{\BBOQ}\APACrefatitle {{Hall magnetic reconnection: Guide field dependence}} {{Hall magnetic reconnection: Guide field dependence}}.{\BBCQ}
\newblock
\APACjournalVolNumPages{Phys. Plasmas}{12}{1}{012322}.
\newblock
\begin{APACrefDOI} \doi{10.1063/1.1834592} \end{APACrefDOI}
\PrintBackRefs{\CurrentBib}

\bibitem [\protect \citeauthoryear {%
{Liu}%
\ \protect \BOthers {.}}{%
{Liu}%
\ \protect \BOthers {.}}{%
{\protect \APACyear {2022}}%
}]{%
Liu_2022}
\APACinsertmetastar {%
Liu_2022}%
\begin{APACrefauthors}%
{Liu}, Y\BHBI H.%
, {Cassak}, P.%
, {Li}, X.%
, {Hesse}, M.%
, {Lin}, S\BHBI C.%
\BCBL {}\ \BBA {} {Genestreti}, K.%
\end{APACrefauthors}%
\unskip\
\newblock
\APACrefYearMonthDay{2022}{{\APACmonth{12}}}{}.
\newblock
{\BBOQ}\APACrefatitle {{First-principles theory of the rate of magnetic reconnection in magnetospheric and solar plasmas}} {{First-principles theory of the rate of magnetic reconnection in magnetospheric and solar plasmas}}.{\BBCQ}
\newblock
\APACjournalVolNumPages{Communications Physics}{5}{1}{97}.
\newblock
\begin{APACrefDOI} \doi{10.1038/s42005-022-00854-x} \end{APACrefDOI}
\PrintBackRefs{\CurrentBib}

\bibitem [\protect \citeauthoryear {%
{Liu}%
, {Daughton}%
, {Karimabadi}%
, {Li}%
\BCBL {}\ \BBA {} {Peter Gary}%
}{%
{Liu}%
\ \protect \BOthers {.}}{%
{\protect \APACyear {2014}}%
}]{%
Liu_2014_dispersive_waves}
\APACinsertmetastar {%
Liu_2014_dispersive_waves}%
\begin{APACrefauthors}%
{Liu}, Y\BHBI H.%
, {Daughton}, W.%
, {Karimabadi}, H.%
, {Li}, H.%
\BCBL {}\ \BBA {} {Peter Gary}, S.%
\end{APACrefauthors}%
\unskip\
\newblock
\APACrefYearMonthDay{2014}{{\APACmonth{02}}}{}.
\newblock
{\BBOQ}\APACrefatitle {{Do dispersive waves play a role in collisionless magnetic reconnection?}} {{Do dispersive waves play a role in collisionless magnetic reconnection?}}{\BBCQ}
\newblock
\APACjournalVolNumPages{Phys. Plasmas}{21}{2}{022113}.
\newblock
\begin{APACrefDOI} \doi{10.1063/1.4865579} \end{APACrefDOI}
\PrintBackRefs{\CurrentBib}

\bibitem [\protect \citeauthoryear {%
Liu%
\ \protect \BOthers {.}}{%
Liu%
\ \protect \BOthers {.}}{%
{\protect \APACyear {2017}}%
}]{%
Liu_2017}
\APACinsertmetastar {%
Liu_2017}%
\begin{APACrefauthors}%
Liu, Y\BHBI H.%
, Hesse, M.%
, Guo, F.%
, Daughton, W.%
, Li, H.%
, Cassak, P\BPBI A.%
\BCBL {}\ \BBA {} Shay, M\BPBI A.%
\end{APACrefauthors}%
\unskip\
\newblock
\APACrefYearMonthDay{2017}{Feb}{}.
\newblock
{\BBOQ}\APACrefatitle {Why does Steady-State Magnetic Reconnection have a Maximum Local Rate of Order 0.1?} {Why does steady-state magnetic reconnection have a maximum local rate of order 0.1?}{\BBCQ}
\newblock
\APACjournalVolNumPages{Phys. Rev. Lett.}{118}{}{085101}.
\newblock
\begin{APACrefDOI} \doi{10.1103/PhysRevLett.118.085101} \end{APACrefDOI}
\PrintBackRefs{\CurrentBib}

\bibitem [\protect \citeauthoryear {%
Liu%
, Hesse%
, Li%
, Kuznetsova%
\BCBL {}\ \BBA {} Le%
}{%
Liu%
\ \protect \BOthers {.}}{%
{\protect \APACyear {2018}}%
}]{%
liu2018}
\APACinsertmetastar {%
liu2018}%
\begin{APACrefauthors}%
Liu, Y\BHBI H.%
, Hesse, M.%
, Li, T\BPBI C.%
, Kuznetsova, M.%
\BCBL {}\ \BBA {} Le, A.%
\end{APACrefauthors}%
\unskip\
\newblock
\APACrefYearMonthDay{2018}{}{}.
\newblock
{\BBOQ}\APACrefatitle {Orientation and Stability of Asymmetric Magnetic Reconnection X Line} {Orientation and stability of asymmetric magnetic reconnection x line}.{\BBCQ}
\newblock
\APACjournalVolNumPages{J. Geophys. Res. Space Phys.}{123}{6}{4908-4920}.
\newblock
\begin{APACrefDOI} \doi{https://doi.org/10.1029/2018JA025410} \end{APACrefDOI}
\PrintBackRefs{\CurrentBib}

\bibitem [\protect \citeauthoryear {%
Liu%
\ \protect \BOthers {.}}{%
Liu%
\ \protect \BOthers {.}}{%
{\protect \APACyear {2025}}%
}]{%
Liu2025}
\APACinsertmetastar {%
Liu2025}%
\begin{APACrefauthors}%
Liu, Y\BHBI H.%
, Pyakurel, P.%
, Li, X.%
, Hesse, M.%
, Bessho, N.%
, Genestreti, K.%
\BCBL {}\ \BBA {} Thapa, S\BPBI B.%
\end{APACrefauthors}%
\unskip\
\newblock
\APACrefYearMonthDay{2025}{}{}.
\newblock
{\BBOQ}\APACrefatitle {An analytical model of ``Electron-Only''magnetic reconnection rates} {An analytical model of ``electron-only''magnetic reconnection rates}.{\BBCQ}
\newblock
\APACjournalVolNumPages{Communications Physics}{8}{1}{128}.
\PrintBackRefs{\CurrentBib}

\bibitem [\protect \citeauthoryear {%
McPherron%
}{%
McPherron%
}{%
{\protect \APACyear {1979}}%
}]{%
McPherron1979}
\APACinsertmetastar {%
McPherron1979}%
\begin{APACrefauthors}%
McPherron, R\BPBI L.%
\end{APACrefauthors}%
\unskip\
\newblock
\APACrefYearMonthDay{1979}{}{}.
\newblock
{\BBOQ}\APACrefatitle {Magnetospheric substorms} {Magnetospheric substorms}.{\BBCQ}
\newblock
\APACjournalVolNumPages{Reviews of Geophysics}{17}{4}{657-681}.
\newblock
\begin{APACrefDOI} \doi{https://doi.org/10.1029/RG017i004p00657} \end{APACrefDOI}
\PrintBackRefs{\CurrentBib}

\bibitem [\protect \citeauthoryear {%
{Nakamura}%
\ \protect \BOthers {.}}{%
{Nakamura}%
\ \protect \BOthers {.}}{%
{\protect \APACyear {2019}}%
}]{%
RNakamura2019}
\APACinsertmetastar {%
RNakamura2019}%
\begin{APACrefauthors}%
{Nakamura}, R.%
, {Genestreti}, K\BPBI J.%
, {Nakamura}, T.%
, {Baumjohann}, W.%
, {Varsani}, A.%
, {Nagai}, T.%
\BDBL {}{Torbert}, R\BPBI B.%
\end{APACrefauthors}%
\unskip\
\newblock
\APACrefYearMonthDay{2019}{{\APACmonth{02}}}{}.
\newblock
{\BBOQ}\APACrefatitle {{Structure of the Current Sheet in the 11 July 2017 Electron Diffusion Region Event}} {{Structure of the Current Sheet in the 11 July 2017 Electron Diffusion Region Event}}.{\BBCQ}
\newblock
\APACjournalVolNumPages{J. Geophys. Res. Space Phys}{124}{2}{1173-1186}.
\newblock
\begin{APACrefDOI} \doi{10.1029/2018JA026028} \end{APACrefDOI}
\PrintBackRefs{\CurrentBib}

\bibitem [\protect \citeauthoryear {%
Nakamura%
\ \protect \BOthers {.}}{%
Nakamura%
\ \protect \BOthers {.}}{%
{\protect \APACyear {2018}}%
}]{%
TKM_Nakamura2018}
\APACinsertmetastar {%
TKM_Nakamura2018}%
\begin{APACrefauthors}%
Nakamura, T\BPBI K\BPBI M.%
, Genestreti, K\BPBI J.%
, Liu, Y\BHBI H.%
, Nakamura, R.%
, Teh, W\BHBI L.%
, Hasegawa, H.%
\BDBL {}Giles, B\BPBI L.%
\end{APACrefauthors}%
\unskip\
\newblock
\APACrefYearMonthDay{2018}{}{}.
\newblock
{\BBOQ}\APACrefatitle {Measurement of the Magnetic Reconnection Rate in the Earth's Magnetotail} {Measurement of the magnetic reconnection rate in the earth's magnetotail}.{\BBCQ}
\newblock
\APACjournalVolNumPages{J. Geophys. Res. Space Phys.}{123}{11}{9150-9168}.
\newblock
\begin{APACrefDOI} \doi{https://doi.org/10.1029/2018JA025713} \end{APACrefDOI}
\PrintBackRefs{\CurrentBib}

\bibitem [\protect \citeauthoryear {%
{Phan}%
\ \protect \BOthers {.}}{%
{Phan}%
\ \protect \BOthers {.}}{%
{\protect \APACyear {2018}}%
}]{%
Phan2018}
\APACinsertmetastar {%
Phan2018}%
\begin{APACrefauthors}%
{Phan}, T\BPBI D.%
, {Eastwood}, J\BPBI P.%
, {Shay}, M\BPBI A.%
, {Drake}, J\BPBI F.%
, {Sonnerup}, B\BPBI U\BPBI {\"O}.%
, {Fujimoto}, M.%
\BDBL {}{Magnes}, W.%
\end{APACrefauthors}%
\unskip\
\newblock
\APACrefYearMonthDay{2018}{{\APACmonth{05}}}{}.
\newblock
{\BBOQ}\APACrefatitle {{Electron magnetic reconnection without ion coupling in Earth's turbulent magnetosheath}} {{Electron magnetic reconnection without ion coupling in Earth's turbulent magnetosheath}}.{\BBCQ}
\newblock
\APACjournalVolNumPages{Nature}{557}{7704}{202-206}.
\newblock
\begin{APACrefDOI} \doi{10.1038/s41586-018-0091-5} \end{APACrefDOI}
\PrintBackRefs{\CurrentBib}

\bibitem [\protect \citeauthoryear {%
Pritchard%
\ \protect \BOthers {.}}{%
Pritchard%
\ \protect \BOthers {.}}{%
{\protect \APACyear {2023}}%
}]{%
Pritchard2023}
\APACinsertmetastar {%
Pritchard2023}%
\begin{APACrefauthors}%
Pritchard, K\BPBI R.%
, Burch, J\BPBI L.%
, Fuselier, S\BPBI A.%
, Genestreti, K\BPBI J.%
, Denton, R\BPBI E.%
, Webster, J\BPBI M.%
\BCBL {}\ \BBA {} Broll, J\BPBI M.%
\end{APACrefauthors}%
\unskip\
\newblock
\APACrefYearMonthDay{2023}{}{}.
\newblock
{\BBOQ}\APACrefatitle {Reconnection Rates at the Earth's Magnetopause and in the Magnetosheath} {Reconnection rates at the earth's magnetopause and in the magnetosheath}.{\BBCQ}
\newblock
\APACjournalVolNumPages{J. Geophys. Res. Space Phys.}{128}{9}{e2023JA031475}.
\newblock
\begin{APACrefDOI} \doi{https://doi.org/10.1029/2023JA031475} \end{APACrefDOI}
\PrintBackRefs{\CurrentBib}

\bibitem [\protect \citeauthoryear {%
{Pritchard}%
\ \protect \BOthers {.}}{%
{Pritchard}%
\ \protect \BOthers {.}}{%
{\protect \APACyear {2019}}%
}]{%
Pritchard2019}
\APACinsertmetastar {%
Pritchard2019}%
\begin{APACrefauthors}%
{Pritchard}, K\BPBI R.%
, {Burch}, J\BPBI L.%
, {Fuselier}, S\BPBI A.%
, {Webster}, J\BPBI M.%
, {Torbert}, R\BPBI B.%
, {Argall}, M\BPBI R.%
\BDBL {}{Strangeway}, R\BPBI J.%
\end{APACrefauthors}%
\unskip\
\newblock
\APACrefYearMonthDay{2019}{{\APACmonth{09}}}{}.
\newblock
{\BBOQ}\APACrefatitle {{Energy Conversion and Electron Acceleration in the Magnetopause Reconnection Diffusion Region}} {{Energy Conversion and Electron Acceleration in the Magnetopause Reconnection Diffusion Region}}.{\BBCQ}
\newblock
\APACjournalVolNumPages{Geophysical. Res. Lett.}{46}{10274}{10,274-10,282}.
\newblock
\begin{APACrefDOI} \doi{10.1029/2019GL084636} \end{APACrefDOI}
\PrintBackRefs{\CurrentBib}

\bibitem [\protect \citeauthoryear {%
Pritchett%
}{%
Pritchett%
}{%
{\protect \APACyear {2001}}%
}]{%
Pritchett2001}
\APACinsertmetastar {%
Pritchett2001}%
\begin{APACrefauthors}%
Pritchett, P\BPBI L.%
\end{APACrefauthors}%
\unskip\
\newblock
\APACrefYearMonthDay{2001}{}{}.
\newblock
{\BBOQ}\APACrefatitle {Geospace Environment Modeling magnetic reconnection challenge: Simulations with a full particle electromagnetic code} {Geospace environment modeling magnetic reconnection challenge: Simulations with a full particle electromagnetic code}.{\BBCQ}
\newblock
\APACjournalVolNumPages{J. Geophys. Res. Space Phys.}{106}{A3}{3783-3798}.
\newblock
\begin{APACrefDOI} \doi{https://doi.org/10.1029/1999JA001006} \end{APACrefDOI}
\PrintBackRefs{\CurrentBib}

\bibitem [\protect \citeauthoryear {%
Pucci%
\ \protect \BOthers {.}}{%
Pucci%
\ \protect \BOthers {.}}{%
{\protect \APACyear {2018}}%
}]{%
Pucci2018}
\APACinsertmetastar {%
Pucci2018}%
\begin{APACrefauthors}%
Pucci, F.%
, Usami, S.%
, Ji, H.%
, Guo, X.%
, Horiuchi, R.%
, Okamura, S.%
\BDBL {}Yoo, J.%
\end{APACrefauthors}%
\unskip\
\newblock
\APACrefYearMonthDay{2018}{12}{}.
\newblock
{\BBOQ}\APACrefatitle {Energy transfer and electron energization in collisionless magnetic reconnection for different guide-field intensities} {Energy transfer and electron energization in collisionless magnetic reconnection for different guide-field intensities}.{\BBCQ}
\newblock
\APACjournalVolNumPages{Phys. Plasmas}{25}{12}{122111}.
\newblock
\begin{APACrefDOI} \doi{10.1063/1.5050992} \end{APACrefDOI}
\PrintBackRefs{\CurrentBib}

\bibitem [\protect \citeauthoryear {%
Rezeau%
, Belmont%
, Manuzzo%
, Aunai%
\BCBL {}\ \BBA {} Dargent%
}{%
Rezeau%
\ \protect \BOthers {.}}{%
{\protect \APACyear {2018}}%
}]{%
rezeau2018}
\APACinsertmetastar {%
rezeau2018}%
\begin{APACrefauthors}%
Rezeau, L.%
, Belmont, G.%
, Manuzzo, R.%
, Aunai, N.%
\BCBL {}\ \BBA {} Dargent, J.%
\end{APACrefauthors}%
\unskip\
\newblock
\APACrefYearMonthDay{2018}{}{}.
\newblock
{\BBOQ}\APACrefatitle {Analyzing the Magnetopause Internal Structure: New Possibilities Offered by MMS Tested in a Case Study} {Analyzing the magnetopause internal structure: New possibilities offered by mms tested in a case study}.{\BBCQ}
\newblock
\APACjournalVolNumPages{J. Geophys. Res. Space Phys.}{123}{1}{227-241}.
\newblock
\begin{APACrefDOI} \doi{https://doi.org/10.1002/2017JA024526} \end{APACrefDOI}
\PrintBackRefs{\CurrentBib}

\bibitem [\protect \citeauthoryear {%
{Shay}%
\ \BBA {} {Drake}%
}{%
{Shay}%
\ \BBA {} {Drake}%
}{%
{\protect \APACyear {1998}}%
}]{%
shay1998}
\APACinsertmetastar {%
shay1998}%
\begin{APACrefauthors}%
{Shay}, M\BPBI A.%
\BCBT {}\ \BBA {} {Drake}, J\BPBI F.%
\end{APACrefauthors}%
\unskip\
\newblock
\APACrefYearMonthDay{1998}{}{}.
\newblock
{\BBOQ}\APACrefatitle {{The role of electron dissipation on the rate of collisionless magnetic reconnection}} {{The role of electron dissipation on the rate of collisionless magnetic reconnection}}.{\BBCQ}
\newblock
\APACjournalVolNumPages{Geophysical. Res. Lett.}{25}{20}{3759-3762}.
\newblock
\begin{APACrefDOI} \doi{10.1029/1998GL900036} \end{APACrefDOI}
\PrintBackRefs{\CurrentBib}

\bibitem [\protect \citeauthoryear {%
Shi%
\ \protect \BOthers {.}}{%
Shi%
\ \protect \BOthers {.}}{%
{\protect \APACyear {2023}}%
}]{%
PeiyunShi2023}
\APACinsertmetastar {%
PeiyunShi2023}%
\begin{APACrefauthors}%
Shi, P.%
, Scime, E\BPBI E.%
, Barbhuiya, M\BPBI H.%
, Cassak, P\BPBI A.%
, Adhikari, S.%
, Swisdak, M.%
\BCBL {}\ \BBA {} Stawarz, J\BPBI E.%
\end{APACrefauthors}%
\unskip\
\newblock
\APACrefYearMonthDay{2023}{Oct}{}.
\newblock
{\BBOQ}\APACrefatitle {Using Direct Laboratory Measurements of Electron Temperature Anisotropy to Identify the Heating Mechanism in Electron-Only Guide Field Magnetic Reconnection} {Using direct laboratory measurements of electron temperature anisotropy to identify the heating mechanism in electron-only guide field magnetic reconnection}.{\BBCQ}
\newblock
\APACjournalVolNumPages{Phys. Rev. Lett.}{131}{}{155101}.
\newblock
\begin{APACrefDOI} \doi{10.1103/PhysRevLett.131.155101} \end{APACrefDOI}
\PrintBackRefs{\CurrentBib}

\bibitem [\protect \citeauthoryear {%
{Shi}%
\ \protect \BOthers {.}}{%
{Shi}%
\ \protect \BOthers {.}}{%
{\protect \APACyear {2005}}%
}]{%
Shi_2005}
\APACinsertmetastar {%
Shi_2005}%
\begin{APACrefauthors}%
{Shi}, Q\BPBI Q.%
, {Shen}, C.%
, {Pu}, Z\BPBI Y.%
, {Dunlop}, M\BPBI W.%
, {Zong}, Q\BPBI G.%
, {Zhang}, H.%
\BDBL {}{Balogh}, A.%
\end{APACrefauthors}%
\unskip\
\newblock
\APACrefYearMonthDay{2005}{{\APACmonth{06}}}{}.
\newblock
{\BBOQ}\APACrefatitle {{Dimensional analysis of observed structures using multipoint magnetic field measurements: Application to Cluster}} {{Dimensional analysis of observed structures using multipoint magnetic field measurements: Application to Cluster}}.{\BBCQ}
\newblock
\APACjournalVolNumPages{Geophys. Res. Lett.}{32}{12}{L12105}.
\newblock
\begin{APACrefDOI} \doi{10.1029/2005GL022454} \end{APACrefDOI}
\PrintBackRefs{\CurrentBib}

\bibitem [\protect \citeauthoryear {%
{Shi}%
\ \protect \BOthers {.}}{%
{Shi}%
\ \protect \BOthers {.}}{%
{\protect \APACyear {2019}}%
}]{%
Shi_2019}
\APACinsertmetastar {%
Shi_2019}%
\begin{APACrefauthors}%
{Shi}, Q\BPBI Q.%
, {Tian}, A\BPBI M.%
, {Bai}, S\BPBI C.%
, {Hasegawa}, H.%
, {Degeling}, A\BPBI W.%
, {Pu}, Z\BPBI Y.%
\BDBL {}{Liu}, Z\BPBI Q.%
\end{APACrefauthors}%
\unskip\
\newblock
\APACrefYearMonthDay{2019}{}{}.
\newblock
{\BBOQ}\APACrefatitle {{Dimensionality, Coordinate System and Reference Frame for Analysis of In-Situ Space Plasma and Field Data}} {{Dimensionality, Coordinate System and Reference Frame for Analysis of In-Situ Space Plasma and Field Data}}.{\BBCQ}
\newblock
\APACjournalVolNumPages{Space Sci. Rev.}{215}{4}{35}.
\newblock
\begin{APACrefDOI} \doi{10.1007/s11214-019-0601-2} \end{APACrefDOI}
\PrintBackRefs{\CurrentBib}

\bibitem [\protect \citeauthoryear {%
{Stechow}%
\ \protect \BOthers {.}}{%
{Stechow}%
\ \protect \BOthers {.}}{%
{\protect \APACyear {2018}}%
}]{%
Stechow2018}
\APACinsertmetastar {%
Stechow2018}%
\begin{APACrefauthors}%
{Stechow}, A\BPBI v.%
, {Fox}, W.%
, {Jara-Almonte}, J.%
, {Yoo}, J.%
, {Ji}, H.%
\BCBL {}\ \BBA {} {Yamada}, M.%
\end{APACrefauthors}%
\unskip\
\newblock
\APACrefYearMonthDay{2018}{{\APACmonth{05}}}{}.
\newblock
{\BBOQ}\APACrefatitle {{Electromagnetic fluctuations during guide field reconnection in a laboratory plasma}} {{Electromagnetic fluctuations during guide field reconnection in a laboratory plasma}}.{\BBCQ}
\newblock
\APACjournalVolNumPages{Phys. Plasmas}{25}{5}{052120}.
\newblock
\begin{APACrefDOI} \doi{10.1063/1.5025827} \end{APACrefDOI}
\PrintBackRefs{\CurrentBib}

\bibitem [\protect \citeauthoryear {%
{Swisdak}%
}{%
{Swisdak}%
}{%
{\protect \APACyear {2016}}%
}]{%
Swisdak2016}
\APACinsertmetastar {%
Swisdak2016}%
\begin{APACrefauthors}%
{Swisdak}, M.%
\end{APACrefauthors}%
\unskip\
\newblock
\APACrefYearMonthDay{2016}{{\APACmonth{01}}}{}.
\newblock
{\BBOQ}\APACrefatitle {{Quantifying gyrotropy in magnetic reconnection}} {{Quantifying gyrotropy in magnetic reconnection}}.{\BBCQ}
\newblock
\APACjournalVolNumPages{Geophysical. Res. Lett.}{43}{1}{43-49}.
\newblock
\begin{APACrefDOI} \doi{10.1002/2015GL066980} \end{APACrefDOI}
\PrintBackRefs{\CurrentBib}

\bibitem [\protect \citeauthoryear {%
{Tharp}%
\ \protect \BOthers {.}}{%
{Tharp}%
\ \protect \BOthers {.}}{%
{\protect \APACyear {2013}}%
}]{%
Tharp2013}
\APACinsertmetastar {%
Tharp2013}%
\begin{APACrefauthors}%
{Tharp}, T\BPBI D.%
, {Yamada}, M.%
, {Ji}, H.%
, {Lawrence}, E.%
, {Dorfman}, S.%
, {Myers}, C.%
\BDBL {}{Bhattacharjee}, A.%
\end{APACrefauthors}%
\unskip\
\newblock
\APACrefYearMonthDay{2013}{{\APACmonth{05}}}{}.
\newblock
{\BBOQ}\APACrefatitle {{Study of the effects of guide field on Hall reconnection}} {{Study of the effects of guide field on Hall reconnection}}.{\BBCQ}
\newblock
\APACjournalVolNumPages{Phys. Plasmas}{20}{5}{055705}.
\newblock
\begin{APACrefDOI} \doi{10.1063/1.4805244} \end{APACrefDOI}
\PrintBackRefs{\CurrentBib}

\bibitem [\protect \citeauthoryear {%
{Torbert}%
\ \protect \BOthers {.}}{%
{Torbert}%
\ \protect \BOthers {.}}{%
{\protect \APACyear {2018}}%
}]{%
torbert_science2018}
\APACinsertmetastar {%
torbert_science2018}%
\begin{APACrefauthors}%
{Torbert}, R\BPBI B.%
, {Burch}, J\BPBI L.%
, {Phan}, T\BPBI D.%
, {Hesse}, M.%
, {Argall}, M\BPBI R.%
, {Shuster}, J.%
\BDBL {}{Saito}, Y.%
\end{APACrefauthors}%
\unskip\
\newblock
\APACrefYearMonthDay{2018}{{\APACmonth{12}}}{}.
\newblock
{\BBOQ}\APACrefatitle {{Electron-scale dynamics of the diffusion region during symmetric magnetic reconnection in space}} {{Electron-scale dynamics of the diffusion region during symmetric magnetic reconnection in space}}.{\BBCQ}
\newblock
\APACjournalVolNumPages{Science}{362}{6421}{1391-1395}.
\newblock
\begin{APACrefDOI} \doi{10.1126/science.aat2998} \end{APACrefDOI}
\PrintBackRefs{\CurrentBib}

\bibitem [\protect \citeauthoryear {%
{Torbert}%
\ \protect \BOthers {.}}{%
{Torbert}%
\ \protect \BOthers {.}}{%
{\protect \APACyear {2016}}%
}]{%
mms_fields_torbert2016}
\APACinsertmetastar {%
mms_fields_torbert2016}%
\begin{APACrefauthors}%
{Torbert}, R\BPBI B.%
, {Russell}, C\BPBI T.%
, {Magnes}, W.%
, {Ergun}, R\BPBI E.%
, {Lindqvist}, P\BPBI A.%
, {Le Contel}, O.%
\BDBL {}{Lappalainen}, K.%
\end{APACrefauthors}%
\unskip\
\newblock
\APACrefYearMonthDay{2016}{{\APACmonth{03}}}{}.
\newblock
{\BBOQ}\APACrefatitle {{The FIELDS Instrument Suite on MMS: Scientific Objectives, Measurements, and Data Products}} {{The FIELDS Instrument Suite on MMS: Scientific Objectives, Measurements, and Data Products}}.{\BBCQ}
\newblock
\APACjournalVolNumPages{Space Sci. Rev.}{199}{1-4}{105-135}.
\newblock
\begin{APACrefDOI} \doi{10.1007/s11214-014-0109-8} \end{APACrefDOI}
\PrintBackRefs{\CurrentBib}

\bibitem [\protect \citeauthoryear {%
{Vasyliunas}%
}{%
{Vasyliunas}%
}{%
{\protect \APACyear {1975}}%
}]{%
Vasyliunas1975}
\APACinsertmetastar {%
Vasyliunas1975}%
\begin{APACrefauthors}%
{Vasyliunas}, V\BPBI M.%
\end{APACrefauthors}%
\unskip\
\newblock
\APACrefYearMonthDay{1975}{{\APACmonth{02}}}{}.
\newblock
{\BBOQ}\APACrefatitle {{Theoretical models of magnetic field line merging, 1.}} {{Theoretical models of magnetic field line merging, 1.}}{\BBCQ}
\newblock
\APACjournalVolNumPages{Reviews of Geophysics and Space Physics}{13}{}{303-336}.
\newblock
\begin{APACrefDOI} \doi{10.1029/RG013i001p00303} \end{APACrefDOI}
\PrintBackRefs{\CurrentBib}

\bibitem [\protect \citeauthoryear {%
Vines%
, Fuselier%
, Petrinec%
, Trattner%
\BCBL {}\ \BBA {} Allen%
}{%
Vines%
\ \protect \BOthers {.}}{%
{\protect \APACyear {2017}}%
}]{%
vines2017}
\APACinsertmetastar {%
vines2017}%
\begin{APACrefauthors}%
Vines, S\BPBI K.%
, Fuselier, S\BPBI A.%
, Petrinec, S\BPBI M.%
, Trattner, K\BPBI J.%
\BCBL {}\ \BBA {} Allen, R\BPBI C.%
\end{APACrefauthors}%
\unskip\
\newblock
\APACrefYearMonthDay{2017}{}{}.
\newblock
{\BBOQ}\APACrefatitle {Occurrence frequency and location of magnetic islands at the dayside magnetopause} {Occurrence frequency and location of magnetic islands at the dayside magnetopause}.{\BBCQ}
\newblock
\APACjournalVolNumPages{J. Geophys. Res. Space Phys.}{122}{4}{4138-4155}.
\newblock
\begin{APACrefDOI} \doi{https://doi.org/10.1002/2016JA023524} \end{APACrefDOI}
\PrintBackRefs{\CurrentBib}

\bibitem [\protect \citeauthoryear {%
Wang%
, Kistler%
, Mouikis%
\BCBL {}\ \BBA {} Petrinec%
}{%
Wang%
\ \protect \BOthers {.}}{%
{\protect \APACyear {2015}}%
}]{%
ShangWang2015}
\APACinsertmetastar {%
ShangWang2015}%
\begin{APACrefauthors}%
Wang, S.%
, Kistler, L\BPBI M.%
, Mouikis, C\BPBI G.%
\BCBL {}\ \BBA {} Petrinec, S\BPBI M.%
\end{APACrefauthors}%
\unskip\
\newblock
\APACrefYearMonthDay{2015}{}{}.
\newblock
{\BBOQ}\APACrefatitle {Dependence of the dayside magnetopause reconnection rate on local conditions} {Dependence of the dayside magnetopause reconnection rate on local conditions}.{\BBCQ}
\newblock
\APACjournalVolNumPages{J. Geophys. Res. Space Phys.}{120}{8}{6386-6408}.
\newblock
\begin{APACrefDOI} \doi{https://doi.org/10.1002/2015JA021524} \end{APACrefDOI}
\PrintBackRefs{\CurrentBib}

\bibitem [\protect \citeauthoryear {%
Webster%
\ \protect \BOthers {.}}{%
Webster%
\ \protect \BOthers {.}}{%
{\protect \APACyear {2018}}%
}]{%
Webster2018}
\APACinsertmetastar {%
Webster2018}%
\begin{APACrefauthors}%
Webster, J\BPBI M.%
, Burch, J\BPBI L.%
, Reiff, P\BPBI H.%
, Daou, A\BPBI G.%
, Genestreti, K\BPBI J.%
, Graham, D\BPBI B.%
\BDBL {}Wilder, F.%
\end{APACrefauthors}%
\unskip\
\newblock
\APACrefYearMonthDay{2018}{}{}.
\newblock
{\BBOQ}\APACrefatitle {Magnetospheric Multiscale Dayside Reconnection Electron Diffusion Region Events} {Magnetospheric multiscale dayside reconnection electron diffusion region events}.{\BBCQ}
\newblock
\APACjournalVolNumPages{J. Geophys. Res. Space Phys.}{123}{6}{4858-4878}.
\newblock
\begin{APACrefDOI} \doi{https://doi.org/10.1029/2018JA025245} \end{APACrefDOI}
\PrintBackRefs{\CurrentBib}

\bibitem [\protect \citeauthoryear {%
{Wilder}%
\ \protect \BOthers {.}}{%
{Wilder}%
\ \protect \BOthers {.}}{%
{\protect \APACyear {2018}}%
}]{%
wilder2018}
\APACinsertmetastar {%
wilder2018}%
\begin{APACrefauthors}%
{Wilder}, F\BPBI D.%
, {Ergun}, R\BPBI E.%
, {Burch}, J\BPBI L.%
, {Ahmadi}, N.%
, {Eriksson}, S.%
, {Phan}, T\BPBI D.%
\BDBL {}{Khotyaintsev}, Y\BPBI V.%
\end{APACrefauthors}%
\unskip\
\newblock
\APACrefYearMonthDay{2018}{}{}.
\newblock
{\BBOQ}\APACrefatitle {{The Role of the Parallel Electric Field in Electron-Scale Dissipation at Reconnecting Currents in the Magnetosheath}} {{The Role of the Parallel Electric Field in Electron-Scale Dissipation at Reconnecting Currents in the Magnetosheath}}.{\BBCQ}
\newblock
\APACjournalVolNumPages{J. Geophys. Res. Space Phys}{123}{8}{6533-6547}.
\newblock
\begin{APACrefDOI} \doi{10.1029/2018JA025529} \end{APACrefDOI}
\PrintBackRefs{\CurrentBib}

\bibitem [\protect \citeauthoryear {%
Zenitani%
, Hesse%
, Klimas%
\BCBL {}\ \BBA {} Kuznetsova%
}{%
Zenitani%
\ \protect \BOthers {.}}{%
{\protect \APACyear {2011}}%
}]{%
Zenitani2011}
\APACinsertmetastar {%
Zenitani2011}%
\begin{APACrefauthors}%
Zenitani, S.%
, Hesse, M.%
, Klimas, A.%
\BCBL {}\ \BBA {} Kuznetsova, M.%
\end{APACrefauthors}%
\unskip\
\newblock
\APACrefYearMonthDay{2011}{May}{}.
\newblock
{\BBOQ}\APACrefatitle {New Measure of the Dissipation Region in Collisionless Magnetic Reconnection} {New measure of the dissipation region in collisionless magnetic reconnection}.{\BBCQ}
\newblock
\APACjournalVolNumPages{Phys. Rev. Lett.}{106}{}{195003}.
\newblock
\begin{APACrefDOI} \doi{10.1103/PhysRevLett.106.195003} \end{APACrefDOI}
\PrintBackRefs{\CurrentBib}

\bibitem [\protect \citeauthoryear {%
{Zhou}%
\ \protect \BOthers {.}}{%
{Zhou}%
\ \protect \BOthers {.}}{%
{\protect \APACyear {2019}}%
}]{%
Zhou2019}
\APACinsertmetastar {%
Zhou2019}%
\begin{APACrefauthors}%
{Zhou}, M.%
, {Deng}, X\BPBI H.%
, {Zhong}, Z\BPBI H.%
, {Pang}, Y.%
, {Tang}, R\BPBI X.%
, {El-Alaoui}, M.%
\BDBL {}{Lindqvist}, P\BPBI A.%
\end{APACrefauthors}%
\unskip\
\newblock
\APACrefYearMonthDay{2019}{{\APACmonth{01}}}{}.
\newblock
{\BBOQ}\APACrefatitle {{Observations of an Electron Diffusion Region in Symmetric Reconnection with Weak Guide Field}} {{Observations of an Electron Diffusion Region in Symmetric Reconnection with Weak Guide Field}}.{\BBCQ}
\newblock
\APACjournalVolNumPages{The Astrophysical Journal}{870}{1}{34}.
\newblock
\begin{APACrefDOI} \doi{10.3847/1538-4357/aaf16f} \end{APACrefDOI}
\PrintBackRefs{\CurrentBib}

\bibitem [\protect \citeauthoryear {%
Zweibel%
\ \BBA {} Yamada%
}{%
Zweibel%
\ \BBA {} Yamada%
}{%
{\protect \APACyear {2009}}%
}]{%
Zweibel_Yamada_2009}
\APACinsertmetastar {%
Zweibel_Yamada_2009}%
\begin{APACrefauthors}%
Zweibel, E.%
\BCBT {}\ \BBA {} Yamada, M.%
\end{APACrefauthors}%
\unskip\
\newblock
\APACrefYearMonthDay{2009}{09}{}.
\newblock
{\BBOQ}\APACrefatitle {Magnetic Reconnection in Astrophysical and Laboratory Plasmas} {Magnetic reconnection in astrophysical and laboratory plasmas}.{\BBCQ}
\newblock
\APACjournalVolNumPages{Annu. Rev. Astron. Astrophys.}{47}{}{291-332}.
\newblock
\begin{APACrefDOI} \doi{10.1146/annurev-astro-082708-101726} \end{APACrefDOI}
\PrintBackRefs{\CurrentBib}

\end{thebibliography}

%
%
%
%
%

\end{document}